\documentclass[nojss,nofooter]{jss}

\usepackage[plainruled,noline]{algorithm2e}
\usepackage{xcolor}
\usepackage{thumbpdf,lmodern}
\usepackage{amsmath,amssymb}
\usepackage{framed}
\usepackage{graphicx}
\usepackage{float}
\usepackage[utf8]{inputenc}

\SetKwInput{KwInput}{Input}  
\SetKwInput{KwOutput}{Output} 

\makeatletter
\def\@algocf@capt@plainruled{above}
\renewcommand{\algocf@caption@plainruled}{\box\algocf@capbox}%
\makeatother

\newtheorem{Def}{Definition}
\newtheorem{prop}{Proposition}

\newtheorem{teo}{Theorem}

\author{Izhar Asael Alonzo Matamoros \\  Universidad de Nacional Autónoma de Honduras \\}
\Plainauthor{Izhar Asael Alonzo Matamoros}

\title{\huge An introduction to computational complexity in Markov Chain Monte Carlo methods}
\Plaintitle{An introduction to computational complexity in Markov Chain Monte Carlo methods}
\Shorttitle{An introduction to computational complexity in MCMC methods}

\Abstract{The aim of this work is to give an introduction to the theoretical background and computational complexity of Markov chain Monte Carlo methods. Most of the mathematical results related to the convergence are not found in most of the statistical references, and computational complexity is still an open question for most of the MCMC methods. In this work, we provide a general overview, references, and discussion about all these theoretical subjects.}

\Plainkeywords{Markov Chains, MCMC, Complexity classes, Probabilistic Turing Machine}

\Address{
  Izhar Asael Alonzo Matamoros\\
  Universidad Nacional Autónoma de Honduras\\
  Departmento de Estadística\\
  Escuela de Matemática, Facultad de Ciencias\\
  Blvd Suyapa, Universidad Nacional Autónoma de Honduras\\
  Tegucigalpa, Honduras\\
  E-mail: \email{asael.alonzo@gmail.com}\\
  URL: \url{https://asaelam.wixsite.com/asael697site}
}

\begin{document}

\section{Introduction}

The Markov Chain Monte Carlo methods, are a  set of algorithms used for optimization, integral approximation, dynamic simulation, efficient counting, and sampling. This methods are widely used  in problems applied to physics, Chemistry, statistics, probability, combinatorial, optimization, numerical analysis among others. The MCMC methods are preferred by statisticians and applied mathematician because of its easy implementation, in some cases fast convergences, and numerical stability. But, because of its complex theoretical background and lack of theoretic convergence diagnostic, the MCMC methods sometimes are referred as black boxes for sampling and posterior estimation \cite{Brooks2011}. The aim of this work is to present a simple review of the theoretical background and computational complexity of MCMC methods, in particular of the Metropolis-Hastings algorithm, one of the most popular algorithm to the time.\\

The plan of this work is as follows. In Section 2, a history background of MCMC is presented. In section 3, the preliminary concepts of Markov chain, convergence and Complexity classes are given as the necessary theoretical background. In Section 4, a statistics review of Bayesian and Classical schemes, the computational problems related to statistical inference and the Metropolis-Hasting algorithm are presented. Finally, in section 5, a review that MCMC methods are equivalent to probabilistic Turing machines and their applications to counting problems is given.

\section{A short story of MCMC}

Markov chain Monte Carlo (MCMC) was invented after the Monte Carlo methods in \textit{Los Alamos},  \cite{Metropolis1953} simulated a liquid in equilibrium with its gas phase. The obvious way to find out about the thermodynamic equilibrium is to simulate the dynamics of the system, and let it run until it reaches equilibrium. For it, they  simulate some Markov chains that have the same stationary distribution as the proposed dynamical system.  This methodology is currently known as the Metropolis algorithm and it was heavily used in physic and chemical experiments until it was discovered by statistician in 1990 and used for the posterior distribution approximation in Bayesian inference. On the same years \cite{Sinclair1988} use them to approximate solutions to combinatorial counting and uniform generation problems.\\

Their is a lot of MCMC variants, in 1970 Hastings generalize Metropolis algorithm using asymmetric distributions. In 1984, Geman and Geman introduce a particular case of the Metropolis-Hasting known as the Gibbs Sampler. Other MCMC methodologies are Reversible Jump MCMC, Neuronal Network-HMC, Sequential Monte Carlo, Quasi-Monte Carlo (QMC),and Hamiltonian Monte Carlo (HMC) \cite{DUANE1987216}.

Because of its fast converges, the Hamiltonian or Hybrid Monte Carlo, is widely used for the approximation of high dimensional posterior distribution in probabilistic models \cite{betancourt2017}, increasing the popularity of Bayesian methods in Health-care, social science, and physical applications, see \cite{Stan} and \cite{Paul2017}.

\section{Preliminary concepts}

This section is divided in three parts, the first gives preliminary definitions of Markov chains, and some important properties. In the second part we present some definitions for analyzing the convergence time to the stationary distribution. Finally, on the third part, we provide an overview of complexity classes and Turing machines.

\subsection*{An overview of Markov chains}

In this section we provide definitions of stochastic process, Markovian steps and some other properties need  for designing the Metropolis algorithm. More information see \cite{Levin2006}.

\begin{Def}
	An stochastic process is a sequence of random variables $X = \{X_t | t \in T\}$ where T is the space set usually denoted as time, and the random variables  are defined in the state set $\Omega$. 
\end{Def}

Unlike random samples, stochastic process does not assume independence between variables, and most of them are classified by their dependency structure. One of the simplest stochastic models are markov chains.

\begin{Def}
	Let be $X = \{X_t | t \in T\}$ an stochastic process that satisfies:
	\begin{equation}
	P(X_t|X_{t-1},X_{t-2}\ldots, X_1,X_0) = P(X_t | X_{t-1})
	\end{equation}	
	The X is called a Markov process, and if the process X just takes values in a finite set $B \subset \Omega$, then X is called a Markov Chain.
\end{Def}

Equation (1) is usually denoted as the \textit{Markov Property} and can be easy interpreted as follows:\\

 \textit{A process is Markovian if the present is only defined by the previous time}.\\

Is valid to specify that only discrete time stochastic process are considered ($T \subset \mathbb{Z}$), the principal reason, is because the impossibility of simulate infinite values in a computer. Every value $x \in \Omega$ is denoted by state, and P(x,y) is the probability of the process to move form the state x to y in one time $P(x,y) = P(X_t = y|X_{t-1} = x)$, called the transition probability. Using this notation, $P^n(x,y)$ represents the probability of moving from x  to y in exactly n times.

\begin{Def}
	A Markov process X is called irreducible if for any two states $x,y  \in \Omega$ there exists an integer $k > 0$ such that $P^k(x,y) > 0$.
\end{Def}
This means that is possible to move from any state to any other in a finite amount of time. With this same ideas we can define the periodicity of an state x.

\begin{Def}
	Lets denote $\tau = \{ t >1 | P^t(x,x) > 0 \}$ the set of times for a chain returns to the start position. Then the period of the state x is the greatest common divisor of $\tau$.
\end{Def}

A process is \textit{aperiodic} if all states have period 1, in other case the process is called \textit{periodic}. Some distributions in stochastic process are invariant over time, and are defined as follows.

\begin{Def}
	An stochastic process has a stationary distribution $\pi$ such as 
	$$\pi(X_t) = \pi(X_{m}) \text{ for all }t,m \in T$$.

\end{Def}
For Markov chains where the transition probability accepts a stochastic matrix representation P, the stationary distribution follows the next equation,
$$\pi = \pi P$$
Another important property is the \textit{time reversible} in Markov chains, this property guaranties the existence of an stationary distribution.

\begin{Def}
	An stochastic Markov process is reversible in time if it satisfies:
	$$\pi(X_t =x) = \pi(X_{t-1}= x)$$
	for all $x \in \Omega$ and stationary distribution $\pi$.
\end{Def}

An interesting result of reversible chains is the following proposition
\begin{prop}
	Let be X an irreducible Markov chain  with transition matrix P, if any distribution $\pi$ that satisfies,
	\begin{equation}
		\pi(x)P(x,y) = \pi(y)P(y,x) \text{ for all }  x,y \in \Omega
	\end{equation}
Then $\pi$ is an stationary distribution.
\end{prop}

\begin{dem}
	Sum both sides of the equation $\pi = \pi P$ over all y:
$$\pi P = \sum_{y \in \Omega}\pi(y)P(y,x) =\sum_{y \in \Omega}\pi(x)P(x,y) = \pi(x) = \pi$$
\end{dem}

(2) is usually denoted as the detailed balance equation. Checking (2) is often the simplest way to verify a particular distribution is stationary.\\

The major applications of MCMC methods is that the simulated chain converges to the stationary \textit{target} distribution. Before presenting the methodology itself, some natural questions are proposed:

\begin{itemize}
	\item How to know if the chain converge to the target distribution?
	\item There exist an methodology to prove the chain's convergences?	
	\item How many simulation are need to reach the stationary distribution?
\end{itemize} 

The following definitions and propositions are presented to solve the first two question, the last one is hard problem (\textit{coNP-hard to be exact}), some results and references will be presented later in this work.

\subsection*{Markov chain's mixing time}

For measuring the chains convergence a parameters which measures the time required by a Markov Chain to \textit{reach} the stationary distribution is required \cite{hsu2017mixing}.

\begin{Def}[Total variance distance]
	The total variance distance of two probability distributions f and g in $\Omega$ is defined by:
	$$||f -g||_{TV} = \sup_{A \in \Omega}|f(A) -g(A)|$$
\end{Def}

In terms of the total variance distance we can define the distance between the simulated step and the stationary distribution as follows:
$$d(t) = \sup_{x \in\Omega} ||P(X_t = x|X_{t-1}) -\pi||_{TV}$$

\begin{Def}[Mixing time]
	Lets be X a Markov chain that converges to a stationary distribution $\pi$ the mixing time t is defined by:
	$$t_{mix}(\epsilon) = min\{t |d(t) < \epsilon\} $$
\end{Def}

The mixing time is just the minimum time, the chain needs to be \textit{"close"} to the stationary distribution by a real $epsilon$. Using this previous definitions we can give the following results.

\begin{teo}[Convergence theorem]
	Suppose that a Markov chain is irreducible and aperiodic with stationary distribution $\pi$. Then for every constants $\alpha \in (0,1)$ exist a $C > 0$ such that:
	$$d(t) \leq C\alpha^t$$
\end{teo}

\begin{dem}
	The proof of this theorem is omitted,  \cite{Levin2006} in page 54 provides a full demonstration.
\end{dem}

This results provides a guarantee that the simulated chain eventually will converge to the target distribution.The next theorem has a great impact in statistics applications where the target distribution is the posterior distribution of an unknown quantity and a expected value is the estimated value. 

\begin{teo}[Ergodic Theorem]
	Let g be a real-valued function defined on $\Omega$. If X is an irreducible Markov Chain, then for any starting distribution f
	$$P_f \left[\lim_{t \to \infty}\dfrac{1}{t}\sum_{s=0}^{t-1}g(X_s) = E_\pi(g)\right]  = 1$$
\end{teo}

\begin{dem}
	The proof of this theorem is omitted,  \cite{Levin2006} in page 59 provides a full demonstration.
\end{dem}

The idea of \textit{Ergodic Theorem} for Markov chains is that time averages is equal to the state average. In other words the expected value of the stationary distribution is simply the average of the chain's simulated values. Mixing time will be important in the next sections, the complexity of an MCMC algorithm is measured with the chains mixing time.

\subsection*{An overview of complexity classes}

Let be P the complexity class  that accepts a deterministic Turing machine (TM) in polynomial time in the size of the input. In the same scheme, the NP class, accepts a non deterministic Turing machine (NTM) in polynomial time. The NP class that accepts decision problems can be extended to  $\#P$ class,that consist of all counting problems associated with the decision problems in NP. Furthermore a general definition of $\#P$  is provided as follows.

\begin{Def}
	A problem $\Pi \in \#P$ if there is a nondeterministic polynomial time Turing machine that for any instance I of $\Pi$, has a number of accepting computations exactly equal to the number of distinct solutions to I. And $\Pi$ is a $\#P$-complete if any problem $\Gamma \in \#P$ can be reduced to $\Pi$ by a polynomial time Turing reduction.
\end{Def}

One example of $\#P$-complete problem is the \textit{0,1-Perm} provided by \cite{Valiant}, \textit{determinate the number of perfect matchings in a bipartite graph G}. There is no known efficient deterministic approximation algorithm for any $\#P$-complete problem, but there exists some efficient randomized algorithms that provides good approximation \cite{Jerrum}. \\

\begin{Def}
	A probabilistic Turing machine (PTM) is a kind of Nondeterministic Turing Machine, with coin 
flip choices instead of nondeterministic choices. A coin flip choice is an unbiased random choice between two successor states.
\end{Def}

A PTM follows only one possible branch of a nondeterministic choice, whereas a NTM follows
them all. The probability that M will follow a given computation branch b is,
$$P[b] = 2^{-1k}$$
Where k is the number of coin flip choices on branch b. The probability that M accepts w is,
$$P[\text{PTM accepts w}] =\sum_{b \in A}P(b)$$ 
Where A is the set of all accepting branches. M is said to reject w if and only if it does not accept w
$$P[\text{PTM rejects w}] = 1 - P[\text{M accepts w}].$$

\begin{Def}
	LeT PP the complexity class that accepts a probabilistic Turing machine, such that the following hold when M is run on $w \in \Sigma^*$
	$$ w \in S \Longrightarrow P[\text{PTM accepts w}] > 0.5 $$ 
	$$ w \not\in S \Longrightarrow P[\text{PTM accepts w}] \leq 0.5$$ 
\end{Def}

It can be prove that $NP \subset PP$ (\cite{Daria} provides a full proof), another important class derived from PP s the bounded PP class (BPP) defined as follows.

\begin{Def}
	LeT BPP the complexity class that accepts a probabilistic Turing machine, such that the following hold when M is run on $w \in \Sigma^*$
	$$ w \in S \Longrightarrow P[\text{PTM accepts w}] > 0.5 +\epsilon $$ 
	$$ w \not\in S \Longrightarrow P[\text{PTM accepts w}] \leq 0.5 -\epsilon$$ 
	where $\epsilon$ is a constant $0 < \epsilon < 0.5$
\end{Def}

By definition $BPP \subset PP$, another useful class is the randomized polynomial time RP class, this class differs from BPP in that it has only one-side error.

\begin{Def}
	LeT BPP the complexity class that accepts a probabilistic Turing machine, such that the following hold when M is run on $w \in \Sigma^*$
	$$ w \in S \Longrightarrow P[\text{PTM accepts w}] > \epsilon $$ 
	$$ w \not\in S \Longrightarrow P[\text{PTM accepts w}] = 0$$ 
	where $0 < \epsilon < 1.$	
\end{Def}

coRP is a derived class defined as the complement of RP, some interesting relationship between classes that can be proved directly are,  $RP \subseteq NP$,  $coRP \subseteq coNP$  and $coRP \cup RP \subseteq BPP$. A further discussion is provided by \cite{Daria}.

\section{MCMC and statistical inference}

The basic idea of statistical inference is to estimate important quantities of a given model, using the available data. There are two principal schemes in statistical inference, classic and Bayesian. In the first scheme, some restrictions are considered, like and all the unknown quantities are considered constant real values. In a Bayesian scheme, every unknown value is considerate a random variable.

\subsection*{Elements of inference}

\begin{Def}
	Let $X =\{X_1,X_2,\ldots,X_n\}$ be a set of independent and identically distributed random variables, with a probability distribution P, where P is depending of unknown quantity $\theta$, the X is denoted as a random sample, and $theta$ as an unknown parameter.
\end{Def}

Most of the classical statistical work, estimates $\theta$ using only the available data via the likelihood function, where $\theta$ is a constant real value.

\begin{Def}
	Le be X a random sample, with unknown parameter $\theta$ then the likelihood function $L:\mathbb{R} \rightarrow \mathbb{R}^+$ is defined as follows
	$$L(\theta) = L(\theta;X) = P(X/\theta) = \prod_{i=1}^nf(X_i/\theta)$$
\end{Def} 

The point estimator  is simply a real value used to approximate $\theta$ the most famous method is the Maximum likelihood estimator.

\begin{Def}
	Let $X =\{X_1,X_2,\ldots,X_n\}$ be a random sample with unknown quantity $\theta$, then the maximum likelihood estimator (MLE) of $\theta$, is a real value that maximize the likelihood function
	$$MLE = arg \max_{\theta} \{L(\theta;X)\} $$
\end{Def} 

In a Bayesian inference scheme the parameter $\theta$ is obtained using the available data, and all the external information provided by the expert. The basic idea of Bayesian statistic is to assume that every unknown parameter $\theta$ is a random variable in a probability space $\Theta$, and estimate its conditional distribution using the Bayes theorem as follows,

$$P(\theta/X) = \dfrac{P(X/\theta)P(\theta)}{P(X)}$$
The previous equation can be simplified as follows
$$P(\theta/X) = k P(X/\theta)P(\theta)$$
$$P(\theta/X) \propto P(X/\theta)P(\theta)$$
Where:
\begin{itemize}
	\item $P(\theta/X)$ is the posterior distribution for $\theta$
	\item $P(X/\theta)$ is the likelihood of the model
	\item $P(\theta)$ is the prior distribution for $\theta$
	\item $k^{-1} = P(X)$ is a proportional constant that does not affect directly the model
\end{itemize}

The major problem of Bayesian inference is finding the proportional constant k, that can be computed directly by,
$$P(X) = \int_{\Theta}P(X,\theta)d\theta = \int_{\Theta}P(X/\theta)P(\theta)d\theta$$

For most of the selected prior distributions, finding k is a real hard task making this scheme unpractical for many years until the MCMC methods where discovered by statisticians. For more details see \cite{Miggon} and \cite{degroot19886}. Before introducing the Metropolis Hasting algorithm we present the Bayesian point estimate analogous.

\begin{Def}
	Let $L(\theta,\delta(X))$ be the loss function of choosing  $\theta$ using the decision rule $\delta$, we denote risk by
	 $$R(\delta) = \int_{\Theta}L(\theta,\delta)P(\theta/X)d\theta$$
	 As the expected posterior loss.
\end{Def} 

We say the a decision rule is optimal if it has a minimum risk, as following we define the Bayesian point estimate using two different loss functions.

\begin{prop}
	Let $L_2 = (\delta -\theta)^2$ be the loss associated with the estimation of $\theta$ by $\delta$. Then the estimator of $\theta$ is the posterior expected value $E_{\theta/X}[\theta]$.
\end{prop}

\begin{dem}
	Let be $\delta_2 = E[\theta]$
	$$R(\delta) = E[(\delta -\theta)^2] = E[((\delta -\delta_2) +(\delta_2-\theta))^2]$$
	$$R(\delta) =(\delta -\delta_2)^2 + E[(\delta_2 -\theta)^2] +2(\delta -\delta_2)E[\delta_2 -\Theta]$$
	$$R(\delta) =(\delta -\delta_2)^2 +V(\theta)$$
	Then for any value of $\delta$ different than $\delta_2$
	$$R(\delta_2) = V(\theta) \leq (\delta- \delta_2)^2 + V(\theta) = R(\delta)$$
	Therefore $\delta_2 = E[\theta]$ is the point estimate associated to $L_2$.
\end{dem}

This estimator is currently known as the posterior mean, another important estimator is the maximum a posteriori (MAP) or the GMLE (Generalized Maximum likelihood estimator) cause it maximize the likelihood function penalized by the prior distribution. 

\begin{prop}
	Let $L_\infty = lim_{\epsilon \to 0}I_{|\theta -\delta|}([\epsilon.\infty[)^2$ be the loss associated with the estimation of $\theta$ by $\delta$. Then the estimator of $\theta$ is the posterior mode.
\end{prop}

\begin{dem}
	$$R(\delta) = E[L_\infty] = \lim_{\epsilon \to 0}\left[\int_{-\infty}^{\delta -\epsilon}P(\theta/X)d\theta + \int_{\delta - \epsilon}^{\delta +\epsilon}P(\theta/X)d\theta + 
	\int_{\delta + \epsilon}^{\infty}P(\theta/X)d\theta \right]$$
	$$ R(\delta) = = 1 -\lim_{\epsilon \to 0}P(\delta -\epsilon < \theta < \delta +\epsilon) = P(\delta) = 1 -P(\delta/X)$$
	Thus $R(\delta)$ is minimized when $P(\delta/X)$ is maximized. Hence the mode is the point estimate associated to $L_\infty$.
\end{dem}

\subsection*{Computational complexity of statistical inference}

Lets present the computational complexity for statistical problems. \cite{Arora2012} provides a demonstration that computing the MLE estimator for $\theta$ is NP-Hard, where the problem can be rewritten as follows:\\

\begin{algorithm}[H]
	\SetAlgoLined
	\caption{ML ESTIMATION: $MLE(p,\Theta)$}
	\KwInput{A random sample $X = \{X;1,\ldots,X_n\}$ with distribution P}
	\KwOutput{A parameter $\theta = arg \max_{\theta_0 \in\Theta}L(\theta_0;X) $}
\end{algorithm}

It it is easy to see that estimating the Generalized Maximum Likelihood GMLE or Maximum a posterior is just a constrained by the prior ($P(\theta)$) MLE problem:

$$MAP(\theta) =  arg \max_{\theta_0 \in\Theta}P(\theta_0/X) = k arg \max_{\theta_0 \in\Theta}L(\theta_0;X)P(\theta_0)$$

Then, estimating the Maximum a posteriori  problem can be written as follows: \\

\begin{algorithm}[H]
	\SetAlgoLined
	\caption{MAP ESTIMATION: $MAP(p,\Theta)$}
	\KwInput{A random sample $X = \{X;1,\ldots,X_n\}$ with distribution P}
	\KwOutput{A parameter $\theta = arg \max_{\theta_0 \in\Theta}P(\theta_0/X) $}
\end{algorithm}

\cite{Tosh} proves that the $MAP(p,\Theta)$ problem is reduced in polynomial time to the  $MLE(p,\Theta)$ problem, and therefore, is NP-Hard. In the same work, \cite{Tosh}  shows that approximate sampling problems for finding the sample distribution of an estimator\footnote{Approximate sampling problems are usually treated with resampling methods such as bootstrap and Jacknife algorithms, that are equivalent to a randomized Turing machines}, does not have a polynomial time Turing machine, unless NP = RP.

\subsection*{The Metropolis-Hasting algorithm}

Let present the most common algorithm for Bayesian inference, physics applications and some other theoretical problems. Let be $X = \{X_1,X_2,\ldots,X_n\}$ a random sample with unknown parameter $\theta$, and let $P(\theta)$ a defined prior where:
\begin{itemize}
	\item m is the number of simulations
	\item $f(\theta/X) = P(X/\theta)P(\theta)$ is the non normalized posterior distribution called the proposal distribution.
	\item $J(\theta_t/\theta_{t-1})$ be the Markov chain's jump distribution 
	\item the metropolis-Hastings ratio
	$$r =\dfrac{f(\theta_*/X)J(\theta_{t-1}/\theta_*)}{f(\theta_{t-1}/X)J(\theta_*/\theta_{t-1})}$$
\end{itemize}

The Metropolis-Hastings algorithm goes as follows:\\

	\begin{algorithm}[H]
		\caption{Metropolis-Hastings algorithm}
		\SetAlgoLined
		\KwResult{A sample $\theta_1,\theta_2,\ldots,\theta_n$ of $P(\theta/X)$}
		Draw a random value $\theta_0$ such that $f(\theta_0/X) > 0 $\;
		\For{ t $\in$ 1:2,3,...,m}{
			Draw a candidate value $\theta_*$ from $J(\theta/\theta_{t-1})$\;
			Calculate the metropolis-Hasting ratio r\;
			Set $\alpha = min(r,1)$\;
			Draw a value U from a U(0,1) distribution\;
			\eIf{U $< \alpha$}{
				Set $\theta_t = \theta_*$\;
			}{
				Set $\theta_t =\theta_{t-1} $\;
			}
		}
	\end{algorithm}

Using that $r(\theta_t,\theta_{t-1}) = 1/r(\theta_{t-1},\theta_t)$ is easy to prove the next formula.

\begin{equation}
	f(\theta_t/X)\alpha(\theta_t,\theta_{t-1})r(\theta_t,\theta_{t-1}) = f(\theta_{t-1}/X)\alpha(\theta_{t-1},\theta_t)r(\theta_{t-1},\theta_t)
\end{equation}

The algorithm simulates an irreducible, reversible time Markov chain with transition probability J, and by \textit{proposition 1}, has an stationary distribution $\pi$, dividing equation (3) by 
$$k = P(X) = \int_{\Theta}f(\theta/X)d\theta$$ 
The stationary distribution $\pi$ is $P(\theta/X)$. The simulated Markov chain $\{\theta_i\}_{i=1}^m$ is a sample of the posterior distribution $P(\theta/X)$, and by the \textit{Ergodic theorem}, the posterior mean  is approximated by:
$$E[\theta/X] \approx \dfrac{1}{m}\sum_{k=1}^{m}\theta_k$$

A further discussion and complete proof is provided by \cite{Brooks2011}. There exist a lot of Metropolis-Hastings(MH) algorithm variations,  must of them are improvements or particular cases of the presented algorithm.  The Random-Walk Metropolis is simply use a Random-walk for jump distribution \cite{Sherlock}. \cite{Brooks2011} provides a demonstration that a Gibbs sampler is just a particular case of the MH. The Metropolis-adjusted Langevien algorithm (MALA) \cite{papamarkou2016} and Hamiltonian Monte Carlo (HMC) \cite{betancourt2017} are MH with an optimal adjusted random jump. The No U-turn sampler algorithm (NUTS) proposed by \cite{hoffman14a} and implemented in \cite{Stan} is just an automatically tunned HMC.

\subsection*{Computational complexity of MCMC}

Give a complexity bound of an MCMC methods by counting the number of operations is a difficult task, and most of it is because it depends on the number of estimated parameters $\theta$, the selected proposal distribution f, and the number of simulations m. Another perspective, is to bound the complexity in terms of the chain's mixing time to its stationary distribution. \cite{Nayantara2011} prove that establish if the Markov chain a the time t is close to stationary, is an NP-hard problem. Even so, \cite{roberts2014} using diffusion limits, give a lower bound $O(d)$ for a Metropolis-Hastings to converge to the stationary distribution, where d is the dimension of the parameter space $\Theta$. On the other hand, on cases of large data, \cite{Belloni2011} provides an upper bound $O(d^2)$ , for a MH to converge to its stationary distribution. For MCMC methods with optimal jump adjustment such as MALA, \cite{roberts2014} provide a lower bound $O(d^{1/3})$  to converge, and \cite{papamarkou2016} provides convergence bounds comparation for MALA, HMC, MMALA (Manifold-MALA) and SMMALA (simplified MMALA).\\
 
\section{MCMC and counting problems}

Following the work of \cite{Sinclair1988} we show the MCMC methods are equivalent to probabilistic Turing Machines, and provide an overview, that this methods can be used for efficient approximation of  some  counting problems of the complexity class $\#P$.\\

The basic idea  is to present a fully-polynomial almost uniform sampler (FPAUS) algorithm as a simulated finite Markov chain that converges to a uniform stationary distribution (MCMC method in a finite state $\Omega$). Finally, we show the utility of MCMC for self reducible counting problems that accepts randomized algorithms.

Let R be a counting problem with no exact solution in polynomial time, then, an approximate counting solution might be possible if R accepts a randomized Turing machine as follows:

\begin{Def}[Fully polynomial randomized approximation scheme]
 A randomized approximate counter for a relation R in an finite alphabet $\Sigma$ with  real valued function $$\rho:\mathbb{N} \rightarrow \mathbb{R}^+$$ is a probabilistic Turing machine C  whose output for every $x \in \Sigma^*$ and two parameters $\epsilon> 0$, and $\delta < 1$, is non negative real value random variable $X(x,\epsilon)$ satisfying
	$$P(X(x,\epsilon)\rho^{-1}|x| \leq \#R \leq X(x,\epsilon) \rho|x| ) \geq 1 -\delta$$ 
	The algorithm is fully polynomial in $|x|$ $\epsilon^-{1}$ and $-log(\delta)$.
\end{Def}

The significance of a lower bound $1 - \delta$ in the previous definition lies in the fact the allows a count approximation, so the probability of producing a bad estimation is low in polynomial time. 

\begin{Def}[Fully-polynomial almost uniform sampler]
	An almost uniform sampler for a relation R in an finite alphabet $\Sigma$ is a probabilistic Turing machine C  whose output for every $x \in \Sigma^*$ and a parameter $\epsilon > 0$, is non negative real value random variable $X(x,\epsilon)$ satisfying
	\begin{itemize}
		\item $X(x,\epsilon)$ takes values in a set $R \cup \{y\}$, with $y \not\in \Sigma$ and
		$$R \neq \emptyset \Longrightarrow P(X = y) \leq 1/2 $$
		\item There exist a function $\phi:\Sigma^* \rightarrow ]0,1]$ such that for all $y \in \Sigma^*$
		$$y \not\in R \Longleftrightarrow P(X = y) = 0$$
		$$y \in R \Longleftrightarrow (1+ \epsilon)^{-1}\phi(x,\epsilon) \leq P(X = y) = 0 \leq (1+ \epsilon)\phi(x,\epsilon) $$
	\end{itemize}
    The algorithm is fully polynomial in $|x|$ and $-log(\epsilon)$.
\end{Def}

Is hard to visualize that a FPAUS algorithm is equivalent to a Markov chain. For it, we present the concept of self-reducibility that allows to visualize the problem as a tree T, and then construct a Markov chain with a random path  in T is intuitive.

\begin{Def}
	A relation R is polynomial time self-reducible if:
	\begin{itemize}
		\item There exists a polynomial time computable length function $I_R:\Sigma^* \rightarrow N$ such that $I_R(x) = O(|x|^)$ for some $k > 0$, and $y \in R \Longleftrightarrow |y| = L_R(x)$ for all $x,y \in \Sigma$ 
		\item For all $x \in \Sigma^*$ with $I_R(x) = 0$, then the predicate $\Lambda \in R$ can be tested in polynomial time
		\item There exists polynomial time computable functions $\phi:\Sigma^* \rightarrow \Sigma^*$ and $\sigma: \Sigma^* \rightarrow N$ satisfying:
		$$\sigma(x) = O(log|x|) $$ 
		$$I_R(x) > 0 \text{ if and only if }  \sigma(x) > 0$$
		$$|\phi(x,w)| \leq |x| \text{ for all } x,w \in \Sigma^*$$
		$$I_R\phi(x,w) = max\{I_R(x)-|w|,0\} \text{ for all } x,w \in \Sigma^*$$
	\end{itemize}
\end{Def}

The inductive construction of solutions of a self-reducible relation explicitly in a tree structure. For each $x \in \Sigma^*$ with $R(x) \neq 0$, the tree of derivations T(x) is a rooted tree, which each vertex v bears both a problem instance label and a partial solution label. Then the constructed sampler views the vertices of the tree of derivations as the states of a Markov chain, in which there is a non-zero transition probability between two states if they are adjacent in the tree.\\

Finally,  a result from \cite{Sinclair1988} is presented, it shows that every self-reducible counting problem that accept as FRPAS also accepts a FPAUS. There fore, an approximate solution might be perform simulating a Markov chain with a uniform stationary distribution, in polynomial time.

\begin{teo}
	Let be R self-reducible. If there exists a polynomial-time approximate counter scheme for R, within ratio $1 + O(n^2)$ for some $X \in R$, then there exist a fully polynomial almost uniform sampler for R.
\end{teo}

\begin{dem}
	A fully proof is given by \cite{Sinclair1988} in \textit{theorem 4.4} and \textit{theorem 4.5}.
\end{dem}

\bibliography{refs}

\end{document}